\documentstyle[12pt]{article}
\def\adot{\dot \alpha}

\newfont{\bbbold}{msbm10 scaled \magstep1}
\def\bbC{\mbox{\bbbold C}}

\def\bbM{\mbox{\bbbold M}}

\def\bbF{\mbox{\bbbold F}}
\def\bbG{\mbox{\bbbold G}}

\def\bbT{\mbox{\bbbold T}}
\def\bbZ{\mbox{\bbbold Z}}
\def\bbU{\mbox{\bbbold U}}

\newfont{\goth}{eufm10 scaled \magstep1}

\def\gg{\mbox{\goth g}}

\def\gl{\mbox{\goth l}}

\def\gs{\mbox{\goth s}}
\def\gu{\mbox{\goth u}}

\def\xz{\times}
\def\a{\alpha}

\def\d{\delta}\def\D{\Delta}
\def\e{\epsilon}

\def\l{\lambda}

\def\p{\pi}

\def\th{\theta}

\def\beq{\begin{equation}}\def\eeq{\end{equation}}
\def\beqa{\begin{eqnarray}}\def\eeqa{\end{eqnarray}}
\def\barr{\begin{array}}\def\earr{\end{array}}

\def\sd{\mbox{\rm sdet\,}}

\def\wt{\widetilde}
\def\uA{\underline{A}}

\def\hx{\hat x}
\begin{document}

\begin{titlepage}
\begin{flushright}
King's College KCL-TH-96-17\\
hepth@xxx/9611075\\
\today
\end{flushright}

\bigskip\bigskip\begin{center} {\bf
\Large{Superconformal Invariants and Extended
Supersymmetry}}
\end{center} \vskip 1.0truecm

\centerline{\bf P.S. Howe}
\vskip 5mm
\centerline{and}
\vskip 5mm
\centerline{\bf P.C. West}
\vskip 5mm
\centerline{Department of Mathematics}
\centerline{King's College, London}
\vskip5mm

\bigskip \nopagebreak \begin{abstract}
\noindent
The superconformal invariants in analytic superspace are found.
Superconformal
invariance is shown to imply that the Green's functions of analytic operators
are invariant holomorphic sections of a line bundle on a product of certain
harmonic superspaces. It is argued  that the correlation functions for a class
of sufficiently low dimension gauge invariant
operators in $N=2$ and $N=4$
supersymmetric Yang-Mills theory can be evaluated up to constants.
\end{abstract}

\end{titlepage}

It is well-known that supersymmetric Yang-Mills theories with $N=4$
supersymmetry possess remarkable properties. Several arguments  have
been  advanced which suggest that these theories are finite in
perturbation theory \cite{finite} and  even non-perturbatively. They
are therefore quantum-mechanically  superconformally invariant.
Moreover \cite{sen},
$N=4$ theories have the right structure to  exhibit Olive-Montonen
duality \cite{duality}.

In a recent paper \cite{world}, we argued that one might be able
to  exploit the symmetry  properties of these theories further to
calculate the Green's functions for  certain gauge-invariant operators
and we produced evidence which indicated that  these might be rather
simple.  In this note we give further details on these results. We show
explicitly how to compute the Green's functions, and how the
constraints imposed by the analyticity properties of the theory
severely restrict the possible  functions that can arise. It turns out
that there are no arbitrary functions of  superconformal invariants
that can occur in certain Green's functions and hence  that they are
determined up to constants. Our computations are at the conformal
point, i.e. there is no symmetry breaking, but our results
nevertheless suggest  that the theory away from the conformal point
might also be rather simple. In  addition the results are restricted
to a certain sector in the theory, the so-called  analytic sector, but
we emphasize that this sector is much larger than the  chiral sectors
of $N=1$ or $N=2$ gauge theories. In particular, the operators we
consider include the energy-momentum tensor multiplet, as well as all
other gauge-invariant polynomials of the analytic field strength
multiplet.

We begin with a brief review of the theory which we will then
reformulate in two different, but related, ways. The basic multiplet
consists of six  scalar fields, four left-handed chiral fermions and
the gauge field, all of which fields transform under the adjoint
representation of the gauge group which  we can take to be $SU(M)$ to
be definite. If we replace the gauge field by its  field strength
tensor, the entire multiplet can be assembled into an $N=4$
superfield $W_{ij},\ i,j=1\ldots 4$, which is antisymmetric on its
$SU(4)$  indices and which obeys the following constraints:
\beqa
\nabla_{\a i} W_{jk}&=&\nabla_{\a [i} W_{jk]} \nonumber \\
\nabla_{\adot}^i W_{jk}&=&-{2\over3}\d^i_{[j}\nabla_{\adot}^l W_{k]l}
\eeqa
as well as the reality condition
\beq
\bar{W}^{ij} = {1\over2} \e^{ijkl} W_{kl}
\eeq
where the bar includes hermitian conjugation for the Lie algebra basis
matrices.
The latter equation implies that only one of  the first two is
independent. These equations involve the superspace gauge-covariant
derivative $\nabla_{\a i}$, and  hence are quite complicated to write
out explicitly. Nevertheless, when one  defines the components of
$W_{ij}$ as the fields obtained by successively evaluating spinorial
covariant  derivatives of $W_{ij}$ at $\theta=0$ one finds exactly the
components listed above.

A more useful formulation of $N=4$ Yang-Mills theory uses harmonic
superspace \cite{harm}. It was observed in \cite{hart} that one could
rewrite the above constraints in a simpler  form if one uses an
appropriate harmonic superspace. If we let $M$ denote
$N=4$  superspace, the harmonic superspace we wish to consider is
$M_H=M\times \bbF$ where $\bbF$ is an internal space which we take to
be the Grassmannian of two-planes in
$\bbC^4$, $\bbG_2(4)$. It can be realised as a homogeneous space of $SU(4)$ by
$\bbF
=H\backslash SU(4)$ where $H=S(U(2)\times U(2))$. Following
Gikos\cite{harm}, we sometimes  find
it convenient to work on the space $\hat M_H:=M\times SU(4)$ with the
understanding that all fields will be equivariant with respect to
the left  action of the isotropy subgroup, that is, $F(hu)=M(h)F(u)$,
where $u\in SU(4)$,
$h\in H$, and where $M$ denotes a represention of $H$ acting on the
representation space $V$, say, of $H$ in which $F$ takes its values. A set of
``coordinates'' for $\hat M_H$ is given by $x^{\a\adot},\th^{\a
i},\bar\theta^{\adot}_i$ and $u_I{}^i$, where $u\in SU(4)$ and where the
capital
index is acted on by the isotropy group. We write
$u_I{}^i=(u_r{}^i,u_{r'}{}^i)$
where $r=1,2$ and $r'=3,4$. The inverse of $u$ is denoted
$u_i{}^I=(u_i{}^r,u_i{}^{r'})$. The right invariant vector fields on
$SU(4)$ are
denoted $\tilde{D}_I{}^J$, where $\tilde{D}_I{}^I=0$. They generate the
left action of $SU(4)$ on itself, from which fact one can deduce the
following simple formulae for their action on the $u$'s:
\beqa
\tilde{D}_I{}^J u_K{}^k &=&
\d_K{}^J u_I{}^k -{1\over4} \d_I{}^J u_K{}^k \nonumber \\
\tilde{D}_I{}^J u_k{}^K &=&
-\d_I{}^K u_k{}^J +{1\over4} \d_I{}^J u_k{}^K
\eeqa
For applications to the coset space we are interested in,  we separate
the   derivatives into a set corresponding to the isotropy group,
$\{D_r{}^s,  D_{r'}{}^{s'}, D_o\}$, and a set corresponding to the
coset space,
$\{D_r{}^{s'},
D_{r'}{}^s\}$. The notation here is that $D_r{}^s$ and $D_{r'}{}^{s'}$ are both
traceless (i.e. correspond to $\gs\gu(2)$'s) and the  $\gu(1)$
derivative $D_o$  is normalised so that
\beq
D_o u_r{}^i= {1\over2}u_r{}^i\qquad D_o u_{r'}{}^i=-{1\over2}u_{r'}{}^i
\eeq

The space $\bbF$ is a complex manifold with complex dimension 4; the
derivatives $D_{r'}{}^{s}$ are essentially equivalent to the holomorphic
derivatives on $\bbF$ while the derivatives $D_{r}{}^{s'}$ are their
complex conjugates and correspond to the antiholomorphic derivatives on
$\bbF$.

Returning to Yang-Mills theory, we define the superfield $W$ by
\beq
W:=\e^{rs} u_r{}^i u_s{}^j W_{ij}
\eeq
It has the following properties: firstly, it is covariantly Grassmann-analytic,
($G$-analytic), which means that
\beq
\nabla_{\a r} W=\nabla _{\adot}^{r'} W=0
\eeq
where small internal indices are converted to large ones and hence
to $r,r'$ by  means of $u_I{}^i$ and its inverse; secondly, it is
analytic in its dependence  on the coordinates of $\bbF$,
($\bbF$-analytic),
\beq
D_{r}{}^{s'} W=0;
\eeq
thirdly, it has $U(1)$ charge 1, so that $D_o W=W$, and finally   it is
real with  respect to a certain real structure on $M_H$. The latter is
defined as follows: for any function $F(u)$ on $SU(4)$, we set
$\wt{F}(u):=\overline{F(ku)}$,  where $k$ is the matrix
\beq
\left(\barr{rr}
0 & 1 \\
-1 & 0
\earr\right)
\eeq
written in block two-by-two form. For $W$, the reality constraint
$W=\wt{W}$  is equivalent to the self-duality constraint in ordinary
superspace. We also remark that the converse holds, namely , given
a field $W$ satisfying (6), (7) and $D_oW=W$, $W=\wt{W}$, it can be
written in the form (5) for a $W_{ij}$ satisfying (1) and (2).

The idea is to consider gauge-invariant operators of the form
$A_m:=\mbox{\rm tr}(W^m),\  m=2\ldots M$.
These
have the property that they are ordinarily $G$-analytic, i.e.
\beq
D_{\a r} A_m = D_{\adot}^{r'}A_m =0
\eeq
as well as being analytic on $\bbF$ and real. Clearly  $A_m$ has
$U(1)$ charge
$m$. The most interesting operator in this set is the  first one
$A_2:=T$,  the energy-momentum tensor multiplet. It has 128+128
components including the  conserved currents for all the symmetries of
the superconformal group.

In order to calculate the Green's functions for these operators it is
convenient to reformulate the theory once more. This is most easily
accomplished
by working in complex spacetime, which is not a drawback since one
can easily  re-impose reality or one can take the view that these
functions  should anyway be  defined in certain regions of products of
complex spacetimes according to the  general axioms of quantum field
theory. The complex superspace relevant for us  can be constructed as
a coset space of the complexified superconformal group
$SL(4|4,\bbC)$. This is not a simple group but it is easier to  work
with than  its
simple cousin because it can be written in terms of matrices.  One can
check that  for the cosets of interest only the simple part acts
non-trivially. The space we  are  interested in is the
super-Grassmannian $\bbG_{2|2}(4|4)$, i.e. the space  of
$(2|2)$-planes in $N=4$ supertwistor space $\bbT_4\cong\bbC^{4|4}$.
The  analytic  functions on this space correspond precisely to
analytic continuations of the  type of functions we have introduced
above, i.e. the $G$-analytic and
$\bbF$-analytic functions on harmonic superspace. This superspace
can be presented as a coset space of $SL(4|4)$ \cite{hart}\cite
{world} with isotropy group
$P$ consisting of supermatrices of the form
\beq
\left(\barr{cccc|cccc}
\xz &\xz & & &\xz &\xz& &  \\
\xz &\xz & & &\xz &\xz& &  \\
\xz &\xz &\xz &\xz &\xz &\xz&\xz &\xz  \\
\xz &\xz &\xz &\xz &\xz &\xz&\xz &\xz  \\
\hline
\xz &\xz & & &\xz &\xz& &  \\
\xz &\xz & & &\xz &\xz& &  \\
\xz &\xz &\xz &\xz &\xz &\xz&\xz &\xz  \\
\xz &\xz &\xz &\xz &\xz &\xz&\xz &\xz
\earr\right)
\eeq
where the crosses denote elements which are not necessarily zero. One can
think of the coset itself as corresponding to the blank portions of the
above diagram.
Notice that both the spacetime  and internal parts of the body,
corresponding to the top left and bottom right blocks respectively, have
the same form, namely the coset space  $P_o\backslash SL(4)$ where $P_o$ is
the subgroup of $SL(4)$ consisting of matrices of the form
\beq
\left(\barr{cccc}
\xz&\xz& & \\
\xz&\xz& & \\
\xz&\xz&\xz &\xz \\
\xz&\xz&\xz &\xz
\earr
\right)
\eeq
This is just the space $\bbF=\bbG_2(4)$. Observe that the number of odd
dimensions of the above superspace (corresponding to the off-diagonal
blocks) is 8, half of that of complex super Minkowski space. Therefore one
can think of fields on this space as being fields defined on complex super
Minkowski space times the internal space $\bbF$, and satisfying the
Grassmann analyticity condition (), or at least the complex version of it.

A slightly different description of the superspace $\bbG_{(2|2)}(4|4)$ is
as follows: let $e_{\uA}$ denote a fixed
set of basis elements for $\bbT_4$, where $\uA$ is a superindex taking $(4|4)$
possible values. Any $(2|2)$-plane is specified by a set of basis elements
$f_A$
where $A$ is a superindex taking $(2|2)$ values. One can then write
\beq
f_A=u_A{}^{\uA} e_{\uA}
\eeq
thus obtaining a description of the plane by the $(2|2)\times (4|4)$
supermatrix
$u$, which is required to have super-rank $(2|2)$. The group
$SL(4|4)$\footnote[1]{We suppress the $\bbC$ for all groups from now on.}
acting
to the left on $\bbT_4$ sends a given plane into another one, and this
corresponds
to a right action on the matrix $u$\footnote[2]{Since this action is
transitive, one can recover the coset space description by finding the
isotropy group $P$ which is simply the one given above.}. However, such a
transformation might simply
mix the basis elements of a given plane, so that we
can describe the space of $(2|2)$ planes as the space of matrices of the above
type, which we shall denote by $\bbU$, modulo the left action of $GL(2|2)$.
By exploiting this fact one can show that the
space can be covered by
coordinate patches of the form
\beq
\bbC^{8|8}\ni X \mapsto s(X):= (1\ \ X)
\eeq
where $1$ denotes the $(2|2)\times (2|2)$ unit matrix and $X$ the $(2|2)\times
(2|2)$ matrix of coordinates:
\beq
X=\left( \barr{ll}
x & \l \\
\p & y \earr\right)
\eeq
Here, each entry is a two-by-two matrix, $x$ is the usual spacetime coordinate
in two-component spinor form, $y$ a local coordinate for the internal space
$\bbF(=\bbG_2(4))$ and $\l$ and $\pi$ are the fermionic coordinates, which
correspond to the $G$-analytic projections of $\th$ and $\bar\th$ respectively.
An important point here is that we wish to consider a subspace of the
super-Grassmannian which will be non-compact as far as the spacetime part is
concerned, but compact as far as the internal part is concerned, i.e. the body
will be complex spacetime $\times\ \bbF$. It is well-known that complex
spacetime
can itself be represented as a subset of $\bbG_2(4)$, but not the whole of it
unless
one wishes to consider its compactified form. To summarise, we shall consider a
subspace $\bbM$ of the super-Grassmanian which can be covered by coordinate
patches of
the above form such that the body is the required one. We shall not give the
details of this here as it will suffice, for our present purposes at least, to
consider only one such patch.

The theory is particularly simple in this formulation. The action of the
superconformal group on the coordinates $X$ is easily obtained, by sending
$s(X)$ to $s(X)g$, where $g\in SL(4|4)$ and then making a compensating
transformation with an element $h(X,g)\in GL(2|2)$ to restore the form of the
function $s$. This gives
\beq
s(X')=h(X,g)s(X)g
\eeq
where $X'$ is the transformed coordinate matrix.
Infinitesimally, with $g\simeq 1+\d g$, $h\simeq 1+\d h$, where $\d
g\in\gs\gl(4|4)$ and $\d h\in\gg\gl(2|2)$, and with $X'\simeq X+\d X$, one
finds
\beq
\d X=B + XD +AX +XCX
\eeq
and
\beq
\d h=-(A+XC)
\eeq
where
\beq
\d g=\left(\barr{ll}
-A & B \\
-C & D
\earr\right)
\eeq
A field $f$ on $\bbM$ is equivalent to an equivariant field $F$ on $\bbU$, that
is a field $F(u)$ such that $F(hu)=M(h)F(u)$ where $M(h)$ now denotes a
representation of $GL(2|2)$ acting on the vector space in which $F$ takes its
values. The (local) correspondence is
\beq
f(X)=F(s(X))
\eeq
The left action of the superconformal group on $F$ is given by $F(u)\mapsto
F'(u)=F(ug)$, and this induces the following action on $f$:
\beq
f(X)\mapsto f'(X):=M(h(X,g))f(X')
\eeq
from which one can easily work out the infinitesimal version explicitly. We
shall only be concerned with fields which transform under one-dimensional
representations of $H$, i.e. fields $F$ such that
\beq
F(hu)= (\sd h)^{-q}F(u)
\eeq
where $q\in\bbZ$ is the $U(1)$ charge of the field in the real
version given above. For such fields one has the infinitesimal
transformation in coordinate form,
\beq
\d f=V f + q\D f
\eeq
where
\beq
V=\d X {\partial\over\partial X}
\eeq
and $\D=-{\rm str}\d h$.

Given this formalism we can state the Ward identities for superconformal
invariance for fields of the above type succinctly. A Green's functions of $n$
fields with charges $q_i,i=1\ldots n$ corresponds to a function $F$ on $\bbU^n$
which satisfies the following equation:
\beq
F(h_1 u_1,\ldots ,h_n u_n)=\prod_{i=1}^n (\sd h_i)^{-q_i} F(u_1g, \ldots ,u_n
g)
\label{wi}
\eeq
for all $h_i\in GL(2|2)$ and $g\in SL(4|4)$.

The problem is therefore to solve this equation. We shall do this by first
finding a set of quasi-invariants, i.e. functions on $\bbU^n$ invariant under
$SL(4|4)$ and under $SL(2|2)^n$. Consider first $SL(4|4)$ invariance. The $n\
u$'s can be considered as a set of vectors in $\bbT$ (of both even and odd
type); if we select out two points, 1 and 2, say, we can form the square matrix
$u_{12}$ which we assume to be non-singular:
\beq
u_{12}:=\left( \barr{c}
u_1 \\ u_2 \earr \right)
\eeq
One can then show that the differential equation expressing the invariance
of a function of these variables under the action
of the super Lie algebra $\gs\gl(4|4)$ is solved by the following quantities:
\beqa
s_{12} &:= & \sd u_{12}  \\
K_i &:=& u_i (u^{-1}_{12})^1, \ i=3\ldots n \\
L_i &:=& u_i (u^{-1}_{12})^2, \ i=3\ldots n
\eeqa
where the 1 and 2 superscripts on the inverse matrix denote its projections
onto the corresponding subspaces. Any $\gs\gl(4|4)$ invariant will then be
a given as a function of these, at least when $u_{12}$ is non-singular.
The superdeterminant $s_{12}$ is invariant under $SL_i(2|2)$ for all $i=1\ldots
n$, (where $SL_i(2|2)$ denotes the $SL$ group associated with point $i$),
and so
is already a quasi-invariant. The other quantities transform as
\beqa
K_i&\mapsto& h_i K_i h_1^{-1} \\
L_i&\mapsto& h_i L_i h_2^{-1}
\eeqa
We now consider invariance under $SL_i(2|2)$, for each $i=3,\ldots n$. We can
apply the same procedure for each $i$ successively to obtain the following sets
of variables:
\beqa
k_i&:=& \sd K_i,\ i=3,\ldots n \\
M_i&:=& (K_i)^{-1} L_i, i=3,\ldots n
\eeqa
assuming the $K_i$ to be invertible. The superdeterminants here are again
quasi-invariants while the $M_i$ transform as
\beq
M_i\mapsto h_1 M_i h_2^{-1}
\eeq
We can now solve for $SL_2(2|2)$ invariance by singling out $M_3$, say, to get
\beqa
m_3&:=&\sd M_3 \\
N_i&:=&M_i M_3^{-1}, i=4,\ldots n
\eeqa
Again, $m_3$ is a quasi-invariant while $N_i$ transforms as
\beq
N_i\mapsto h_1 N_i h_1^{-1}
\eeq
Finally we need to find the invariants of the $N_i$ under the above action of
$SL_1(2|2)$. Clearly the supertrace of $N_i$ is invariant, and if we remove
this the remaining matrix is in the adjoint representation of $SL(2|2)$;
the invariants under the adjoint action of this group are then given by the
Casimirs which are the independent supertraces from second to fourth order
in the $N$'s. In other words, one adds to the list of invariants the
supertraces of the $N_i$'s from first to fourth order. It is not difficult
to see that these supertraces are fully invariant, and not
just quasi-invariant.

The procedure can be repeated for any pair of points $(i,j)$ chosen at the
intial step; i.e. by replacing $u_1,\ u_2$ by $u_i,\ u_j$ and then carrying
through the same construction. In this way one
generates a large number of quasi-invariants. However, not all of them are
independent and it is easy to show that one can replace the invariants of type
$k_i$ and $m_3$ by invariants of the first type. Thus one arrives at the
following set of functions: $s_{ij}:=\sd u_{ij}$, for each pair of points, and
supertraces of up to the fourth order of the variables of type $N_i$, which are
defined for each set of four distinct points. The latter are superconformal
invariants while the former transform as
\beq
s_{ij}\mapsto (\sd h_i) (\sd h_j) s_{ij}
\eeq
One can form full invariants from these by taking cross-ratios, for example of
the form
\beq
{s_{12} s_{34}\over s_{13} s_{14}}
\eeq
which is clearly invariant. One thus has two types of invariant: type I, the
super cross ratios, and type II, supertraces of products of the $N$'s up to
the fourth power, and of course any function of these. It is not clear how
many of these are independent, but we believe that any invariant can be
written in terms of them, i.e. they form an overcomplete set.

In order to understand these functions better, it is useful to write them in
the standard local coordinate patch. The basic building block, the inverse
superdeterminant of $u_{12}$, is the two-point
function for an abelian field strength $W$ for which $q=1$; it is
\beq
<W(1),W(2)>:=G_{12}\propto
s_{12}^{-1}={y_{12}^2\over \hx_{12}^2}
\eeq
where
\beq
\hx_{12}=x_{12}-\l_{12}y_{12}^{-1}\p_{12}
\eeq
and $x_{12}=x_1-x_2$, etc. Thus the super cross-ratios  are simply
ratios of ordinary cross-ratios for the $x$'s and $y$'s with
supersymmetry being taken  care of by the hats. In fact, $\hx_{12}$ and
$y_{12}$
are $Q$-supersymmetric variables; one could equally well have chosen
to have worked with the $S$-supersymmetric variables, $x_{12}$ and
$\hat y_{12}=y_{12}-\p_{12}x^{-1}\l_{12}$. The simplest  invariant of
type II, the supertrace of one $N$ matrix associated with four points
1,2,3 and 4, say, can be written in local coordinates as
\beq
{\rm str} N={\rm str}(X_{12}^{-1}X_{23} X_{34}{}^{-1}X_{41})
\eeq
Evaluating this, one finds that the leading term is a linear combination of
$x$ cross ratios minus the same linear combination of $y$ cross ratios, and
the function continues as a power series in ${\l\p\over y}$.

Higher point functions can be parametrised as free $n$-point functions times
invariants, with the external charges taken care of by the  free part.
As in any conformal field theory, there are invariants from 4 points
up, but they are all singular in the $y$ variables. Since the $x$ and
$y$ sectors appear symmetrically in the theory, they would also be
singular in the
$x$'s if it were  not for the fact that these singular points are
excluded from the domains of  definitions of the functions. However,
the internal part of the space really is the entire Grassmannian
$\bbG_2(4)$, and so the  singular internal points  cannot be excluded.
On the other hand, it is easy to see that the Green's functions  must
be perfectly regular in the $y$'s, and this places very strong
constraints on the functions of invariants that can arise.

Since any Green's function transforms as in equation (24),
 we can write it as the product of two point Green's functions to suitable
powers multiplied by a
superconformal  invariant. The latter can only be a function of the type I and
type II invariants we have constructed above. In particular, if  we take all
the operators in $G$ to have the same charge $N$ for simplicity,  the
four-point function can be written in the form
\beq
G=(G_{12})^N (G_{34})^N I
\eeq
where $I$ is a superconformal invariant function of $X_1,\ldots X_4$. Therefore
the analyticity properties $G$ can be enforced if we know the superconformal
invariants. If the function $I$ is composed of only type I invariants,  i.e.
super cross ratios,  then it is easy to convince oneself that analyticity
requires that the Green's function be of free field theory form. For example,
the free Green's function for four energy-momentum tensors $T$  is of the form
\beq
<T(1)T(2)T(3)T(4)>\propto (G_{12})^{2}(G_{34})^{2} + {\rm perms}
\eeq
One finds that the only choices of invariant functions $I$ consistent with
analyticity generate the permuted terms with arbitrary constants.

The situation once we allow for type II invariants is different.
Invariants of this type have, in their $(\l,\p)$-expansions  singular terms of
the form
$(y_{12}^2 y_{34}^2)^{-3}$ and, at first sight,  it may appear that that they
could not contribute to the above four point Green's function. However, it is
possible that such singularities in $y_{ij}$ could be cancelled by zeroes
arising from a suitable combination of type I invariants. Although it is clear
that analyticity places very strong constraints on the form of the
Green's functions, a detailed caculation is required to find if analyticity
is sufficiently strong to  determine the Green's function up to constants.
This calculation is long and complicated, but has been carried out for the
four point Green's functions of  N=2 analytic operators and will be reported on
in detail
elsewhere \cite{hw2}. The result is that such Green's functions can be
completely
determined for operators of charge two and three, but arbitary
functions can occur in the Green's functions of higher charge
operators.
It is likely that  this  calculation can be extended to
higher point Green's functions in
$N=2$  theories with similar conclusions.
One would also expect to find that the Green's functions
in $N=4$ Yang-Mills theory are determined up to constants by
superconformal invariance alone for a class of sufficiently low
dimension analytic operators.

We can view the above discussion from a more abstract point of view.
According to equation (24) any $N$ point function is an invariant holomorphic
section of a line bundle on $N$
copies of analytic superspace, the particular line bundle being determined by
the $U(1)$ weight of the operators involved. A  Green's function will
be determined up to constants if there are only a finite number of such
sections of the approriate bundle. However, we are unaware of any general
theorems
that classify the number of such sections.

In some senses it is to be expected that one cannot determine the Green's
functions for
all operators  since this would implicitly require  a definition of
the theory with an action involving operators of  arbitarily high
dimension. However, the Green's functions could be further restricted by
requiring that
they
satisfy physical properties such as crossing and
unitarity. One may also be able to use the bootstrap programme to
determine higher point Green's functions in terms of lower point
Green's functions.

Finally, we comment on the heuristic argument given at the end of  reference
\cite{ope}. There it was argued that the form of the operator product expansion
for analytic operators involves a finite number of primary fields and one
could, as
a result, hope to solve for  Green's functions in terms of the constants
that appear in the operator product expansion. This argument relies on the
operator product expansion being meaningful, i.e. convergent. While this may
not be the case for operators of high dimension the results given in this paper
suggest that it at least seems to be correct for (analytic) operators of
sufficiently low dimension. We also note that the argument of reference
\cite {ope} uses the bootstrap idea which is not used in the
arguments presented for solvability in this paper.

It would be of interest to extend the analysis outlined here to spontaneously
broken superconformal symmetry. If one could calculate some
Green's functions in this phase one might hope to be able to verify the
predictions of duality directly. Finally, it is also possible to study
anomalous superconformal Ward Identities, for example in $N=2$ theories; this
was done in \cite{n2} where it was used to derive the `Matone Identity'\cite{m}
for the Seiberg-Witten prepotential \cite{sw}. This relation was also derived
using instanton methods in \cite{dorey} In an interesting recent paper
\cite{ma}, it
has been shown that this identity was sufficent to recover the enitire
solution of Seiberg-Witten for the prepotential \cite{sw}

\end{document}